\documentclass[twocolumn,prl,superscriptaddress]{revtex4}

\usepackage{amsmath}
\usepackage{amsthm}
\usepackage{graphicx}
\usepackage{amssymb}
\usepackage{amsfonts}

\newcommand{\ket}[1]{\left|{#1}\right\rangle}
\newcommand{\bra}[1]{\left\langle {#1}\right|}

\begin{document}

\title{Arbitrarily little knowledge can give a quantum advantage for nonlocal tasks}

\author{Jonathan Allcock}
\email[Electronic address: ]{jon.allcock@bristol.ac.uk}
\affiliation{Department of Mathematics, University of Bristol,
University Walk, Bristol BS8 1TW, United Kingdom}

\author{Harry Buhrman}
%\email[Electronic address: ]{buhrman@cwi.nl}
\affiliation{CWI, Science Park 123, 1098 XG Amsterdam, The Netherlands}

\author{Noah Linden}
%\email[Electronic address: ]{n.linden@bristol.ac.uk}
\affiliation{Department of Mathematics, University of Bristol,
University Walk, Bristol BS8 1TW, United Kingdom}

\date{2nd March  2009}

\begin{abstract}

It has previously been shown that quantum nonlocality offers no benefit over
classical correlations for performing a distributed task known as nonlocal computation.
This is where separated parties must compute the value of a function without individually
learning anything about the inputs. We show that giving the parties some knowledge
of the inputs, however small, is sufficient to \lq\lq unlock\rq\rq the power of quantum
mechanics to out-perform classical mechanics. 
%In doing so, we reveal a novel feature of the nonlocality embodied in the well-known nonlocal task of Clauser, Horne, Shimony and Holt.
This role of information held locally gives new insight into the general
question of when quantum nonlocality gives an advantage over
classical physics.  Our results also reveal a novel feature of the nonlocality
embodied in the celebrated task of Clauser, Horne, Shimony and Holt.

\end{abstract}

%\pacs{ }
%\keywords{}

\maketitle

Quantum theory allows for separated bodies to be correlated with one another more strongly than is achievable by any local classical means. Once viewed with suspicion, such correlations are now looked at as a valuable resource which enables certain tasks to be performed - including quantum cryptography \cite{Eke91}, teleportation \cite{BBCJ+93}, dense-coding \cite{BW92} and reducing communication complexity \cite{BC97,Wol02} - which would otherwise be impossible classically. 
%Following added by NL
It is a fundamental question for quantum physics to characterise those
tasks for which quantum resources offer a benefit over classical physics;
this is the subject of this Letter.

Perhaps the most celebrated task for which quantum theory offers an advantage over
classical mechanics is due to Clauser, Horne, Shimony and Holt (CHSH) \cite{CHSH69}. In this task, Alice and Bob are spatially separated, and 
each possess %are each in possessesion of 
a single particle. They are then each sent a uniformly random single input bit ($z_1$ and $z_2$ respectively), the value of which determines which of two measurements (with possible outcomes $0$ or $1$) they must perform on their particles. Conditioned on their measurement outcomes, Alice and Bob then each return a single bit ($a$ and $b$ respectively) with the aim of satisfying the following equation with as high a success probability as possible:
\begin{equation}
    a\oplus b = AND(z_1, z_2) = z_1 z_2.
\end{equation}
Thus, the output bit $c$ of the AND function is the XOR of the two bits $a$ and $b$, i.e. $c=a\oplus b$;
this output is not known to Alice or Bob.
As is well-known, if the particles are classical, the maximum success probability is $ 75\% $,
but quantum correlations between the particles allow a success probability of $ \sim 85\% $ to be achieved.  Thus,
quantum mechanics allows for greater success in solving this nonlocal task with the separated
inputs $z_1$ and $z_2$.  We will return to this fundamental example later, as we will see that previous
work on it has missed an intriguing aspect of the nature of the
inputs and outputs.

The task described above is asymmetric between the inputs and outputs in an important way:
the output $c=a\oplus b$ is not known to Alice or Bob, whereas the inputs are.  This led
the authors of
\cite{LPSW07}, following \cite{BBLM+06}, to consider the situation where the input bits of a function are
also distributed between Alice and Bob in such a way that neither has any knowledge of them:
given an input bit $z_{j}$, Alice receives $x_{j}$ and Bob $y_{j}$ such that $z_{j}=x_{j}\oplus y_{j}$,
and $x_{j}$ and  $y_{j}$ are locally random, being with equal probability 0 or 1. This scenario is known as \textit{nonlocal computation}.  Surprisingly, 
it was found in \cite{LPSW07} that quantum mechanics gives no benefit over classical mechanics
for the nonlocal computation of any function $f(z_1, z_2,...,z_n)$ of $n$ bits. This result is simple and rather general. For example, it is true for any probability distribution
on the input bits $z_{j}$ and it does not matter how many parties the bits are distributed to;
\cite{LPSW07} also describes more general families of tasks for which quantum mechanics offers
no advantage. This
leads to the natural question as to whether this is the generic situation.
Does quantum mechanics typically help for nonlocal tasks or not?

In this Letter we probe this question in the following way.  In nonlocal computation, it is important that Alice and Bob individually know nothing about the inputs $z_{j}$.  Here we consider
that they are allowed a small amount of information about the inputs.   By this we mean, for
example, that rather than being totally uncorrelated with $z_{j}$, we allow $x_{j}$ to have a probability
$p\neq1/2$ of being equal to $z_{j}$.  We will show that for a series of tasks, even if $p$ is
arbitrarily close to (but not equal to) $1/2$ , this small amount of knowledge \lq\lq unlocks\rq\rq the power of quantum
mechanics to out-perform classical mechanics.

The plan of this Letter is as follows.  Firstly, we consider the nonlocal $AND$ of two bits and show
explicitly how if Alice and Bob have an infinitesimal amount of knowledge of the input bits, then there
are quantum strategies that out-perform any classical ones.  We then give a family of functions
with input sizes of increasing length for which the same is true.
At the end of the letter we return to the CHSH task and argue that our results give new insight
into the nonlocality it embodies.

Let us briefly review the concept of the nonlocal computation of a Boolean function $f:\left\{0,1\right\}^{n}\times\left\{0,1\right\}^{n}\rightarrow\left\{0,1\right\}$ from $2n$ bits to a single bit \cite{LPSW07}.  This can be described as a particular kind of \textit{nonlocal XOR game} \cite{CHTW04} $G=(f,\pi)$
between a \textit{verifier} $V$ and two \textit{provers}, Alice and Bob.  In a general XOR game, the verifier selects a pair of $n$-bit strings $x = x_{1}x_{2}\ldots x_{n}$ and $y = y_{1}y_{2}\ldots y_{n}$ %$(x,y)\in\left\{0,1\right\}^{n}\times\left\{0,1\right\}^{n}$
according to some joint probability distribution $\pi(x,y)$, and sends $x$ to Alice and $y$ to Bob.
In response, Alice returns a single bit $a$ to the verifier, and similarly Bob returns a single bit $b$. The verifier deems the computation of $f$ to be successful if $a\oplus b = f(x,y)$, in which case Alice and Bob are said to win the game. Before the game commences, Alice and Bob 
%(who are knowledgeable of $f$) 
(who know $f$)
can meet to agree on a common strategy, but once the game has started they are forbidden from communicating with one another.  The nonlocal computation of $f$ corresponds to an XOR game where the following extra two conditions are imposed. Firstly, $f$ cannot depend on $x$ and $y$ individually, but only on their bitwise XOR:
\begin{equation}
    f(x,y) \equiv f(x\oplus y)\equiv f(z), \label{e:distribute}
\end{equation}
where $z = x\oplus y$ denotes the $n$-bit string $z = z_{1}z_{2}\ldots z_{n}$ where $z_{i} = x_{i}\oplus y_{i}$. The second requirement is that
\begin{equation}
    \pi(x,y) = \frac{1}{2^{n}}\widetilde{p}(x\oplus y), \label{e:noknowledge}
\end{equation}
where $\widetilde{p}(x\oplus y) = \widetilde{p}(z)$ is an arbitrary probability distribution on $z = x \oplus y$.
This ensures that neither Alice nor Bob has any knowledge about the inputs $z_{i}$ of $f$, because regardless of the value of $z_{i}$, Alice and Bob receive bits $x_{i}$ and $y_{i}$ which are uniformly random.%, being with equal probability $0$ or $1$.

Following \cite{CHTW04} we define the \textit{classical value} $\omega_{C}(G)$ of an XOR game $G=(f,\pi)$ to be the maximum probability with which Alice and Bob can win using purely classical (deterministic) strategies. Such a strategy corresponds to Alice and Bob choosing their output bits $a$ and $b$ to be functions of $x$ and $y$ respectively. It can be shown that
$\omega_{C}(G) = \frac{1}{2}\left(1+ \varepsilon_{C}(G)\right)$, where the \textit{classical bias} $\varepsilon_{C}$ is given by
\begin{equation}
    \varepsilon_{C}(G) = \max \sum_{x,y}\pi(x,y)(-1)^{f(x,y)}A_{x}B_{y},\label{e:c_bias}
\end{equation}
and the maximum is taken over all
$A_{x},B_{y}\in\left\{-1,1\right\}$. Note that $\varepsilon_{C}$ is
really twice the real bias of the success probability.  Similarly we define the \textit{quantum value} $\omega_{Q}(G)$ to be the maximum probability of successfully computing $f$ when Alice and Bob utilize quantum strategies. Such strategies correspond to Alice and Bob sharing an entangled state $\ket{\psi}$ and, dependent on $x$ and $y$, performing projective measurements on their respective subsystems corresponding to Hermitian operators $a_{x}$ and $b_{y}$ with eigenvalues $0$ and $1$. They then return their measurement results to the verifier. Analogously to the classical case, $\omega_{Q}$ can be expressed as
    $\omega_{Q}(G) = \frac{1}{2}\left(1+ \varepsilon_{Q}(G)\right)$,
where the \textit{quantum bias} $\varepsilon_{Q}$ has the form
\begin{equation*}
    \varepsilon_{Q}(G) = \max \sum_{x,y}\pi(x,y)(-1)^{f(x,y)}\bra{\psi}A_{x}B_{y}\ket{\psi}\\,
\end{equation*}
and the maximum is taken over all pure states $\ket{\psi}$ and Hermitian operators $A_{x}$ and $B_{y}$ (on Alice and Bob's subsystems respectively) with eigenvalues $\pm 1$.

The surprising result of \cite{LPSW07} is that there is no quantum advantage for the nonlocal computation of any function $f$. That is, 
when conditions \eqref{e:distribute} and \eqref{e:noknowledge} are imposed, $\omega_{C}(G) = \omega_{Q}(G)$, and quantum strategies do not allow the computation to succeed with higher probability than is classically possible. As an example consider the simplest interesting case, which we shall denote by $G_{AND}$. This is the game $G_{AND} = (f,\pi)$ where $f$ is the $2$-bit nonlocal AND function:
\begin{equation}
    f(x\oplus y) = AND(z_{1},z_{2}) = z_{1}z_{2}, \label{e:nland}
\end{equation}
where $z$ is drawn according to the uniform distribution over two bits (i.e. $\pi(x,y) = (1/4)\widetilde{p}(z) = 1/16$). 
%Then, the classical and quantum values of the game correspond to the biases
%\begin{equation*}
%    \begin{split}
%     \varepsilon_{C,Q}(G_{AND}) &=  \max \mathbb{E}\left[\sum_{x,y}\pi(x,y)(-1)^{f(x\oplus y)}A_{x}B_{y}\right]\\
%     =\max\frac{1}{16}&\mathbb{E} \left[A_{00}B_{00} + A_{00}B_{01} + A_{00}B_{10} -        A_{00}B_{11}\right.\\
%                                            &+ A_{01}B_{00} + A_{01}B_{01} - A_{01}B_{10} + A_{01}B_{11}\\
%                                            &+ A_{10}B_{00} - A_{01}B_{01} + A_{10}B_{10} + A_{10}B_{11}\\
%                                            &\left.- A_{11}B_{00} + A_{11}B_{01} + A_{11}B_{10} + A_{11}B_{11}\right],
%     \end{split}
%\end{equation*}
%where the maximum is over  $A_{x}$ and $B_{y}$ and they are taken   either $\pm 1$ valued
%real numbers in the classical case and Hermitian operators with
%eigenvalues $\pm 1$, where in the latter case $A_xB_y$ should be read as
%$\bra{\psi}A_{x}B_{y}\ket{\psi}$ for some state $\ket{\psi}$ which is
%also part of the  maximization.
%As shown in \cite{LPSW07}, this expression has the same maximum value ($1/2$) for both classical and quantum strategies.
As shown in \cite{LPSW07}, the classical and quantum biases for this game are both equal to $1/2$. 
However, suppose we relax condition \eqref{e:noknowledge}, and allow Alice and Bob some local knowledge of the input bits $z_{i}$. For instance, suppose that Alice's bit $x_{1}$ has some probability $p$ of being equal to $z_{1}$, and Bob's bit $y_{2}$ has some probability $q$ of being equal to $z_{2}$. (Note that we still require $z_{i} = x_{i}\oplus y_{i}$.) Without loss of generality let us take $p,q\geq 1/2$. Denoting this new \textit{perturbed game} by $G_{AND}^{p,q}$, the classical and quantum biases can be collectively expressed as:
\begin{widetext}
    \begin{align}\label{e:pq}
\varepsilon_{C,Q}(G_{AND}^{p,q})= \max\frac{1}{4}\mathbb{E} &\left[pqA_{00}B_{00} + pqA_{00}B_{01} + (1-p)qA_{00}B_{10} - (1-p)qA_{00}B_{11}\right.\\
&+ p(1-q)A_{01}B_{00} + p(1-q)A_{01}B_{01}-(1-p)(1-q)A_{01}B_{10}+(1-p)(1-q)A_{01}B_{11}\notag\\
&+  pqA_{10}B_{00}- pqA_{10}B_{01} + (1-p)qA_{10}B_{10} +(1-p)qA_{10}B_{11}   \notag \\
& \left. - p(1-q)A_{11}B_{00} + p(1-q)A_{11}B_{01}+ (1-p)(1-q)A_{11}B_{10} +(1-p)(1-q)A_{11}B_{11} \notag
 \right].
    \end{align}
\end{widetext}
Note that the case $p = q = 1/2$ is equivalent to the original game
$G_{AND}$ and has $\varepsilon_{C}(G_{AND})=1/2$
(c.~f.~equation~\ref{e:region1}). The case $p = q = 1$ corresponds to
the standard CHSH task with  
$\varepsilon_{C}(G_{AND}^{1,1})=1/2$.  A bit of thought shows that
$\varepsilon_{C}(G_{AND}^{p,q})$ remains  1/2, for any  $p$ and
$q$. The reason is that the bias cannot {\em increase} when Alice and
Bob have {\ less} information on  $z_1$ and/or $z_2$.  Any protocol for
$G_{AND}^{p,q}$ $(p,q\geq 1/2)$ with bias $b$ can be used to solve
$G_{AND}^{1,1}$ with the same bias $b$ as follows. 
Alice and Bob use two shared random bits $r_1,r_2$ such that
Pr$[r_1=0]=p$ and Pr$[r_2=0]=q$. Alice uses $x_i\oplus r_i$ as her
 inputs to the protocol for $G_{AND}^{p,q}$ and Bob uses $y_i\oplus r_i$.
It is easy to see that for these new inputs 
Pr$[x_1=z_1]=p$ and Pr$[y_2=z_2]=q$. A standard convexity argument
shows that the above reduction can be massaged into a deterministic protocol.
Hence the bias cannot increase  for $p,q <1$ and has to remain $1/2$.

%HMB changed this part the above argument can be used for more inputs
%as well.
%% To see this consider the terms in \eqref{e:pq} with coefficients $pq$:
%% \begin{equation*}
%% \frac{pq}{4}\mathbb{E}\left[A_{00}B_{00} + A_{00}B_{01} + A_{10}B_{00} - A_{10}B_{01}\right].
%% \end{equation*}
%% This is an expression of the form found in the well-known CHSH inequality \cite{CHSH69}, and its maximum possible value is $pq/2$. Similarly, looking at terms with coefficients $p(1-q)$, $(1-p)q$ and $(1-p)(1-q)$, one can see that
%% \begin{align*}
%% \varepsilon_{C}(G_{AND}^{p,q}) &\leq \frac{1}{2}\left[pq + p(1-q) + (1-p)q + (1-p)(1-q)\right]\\
%%   &= \frac{1}{2}.
%% \end{align*}
%% Indeed, this value can be achieved if Alice and Bob choose the
%% strategy $A_{i} = B_{i} = +1$, regardless of the input that they
%% receive. 

On the other hand, by using quantum strategies Alice and Bob can do better.  It is straightforward (although somewhat lengthy) to show that when $1 \geq(2q)^{-1} >p \geq 1/2$ (call this region 1),
\begin{equation}\label{e:region1}
    \varepsilon_{Q}(G_{AND}^{p,q})\leq \sqrt{q^{2}+(1-q)^{2}}\sqrt{p^{2}+(1-p)^{2}},
\end{equation}
and when $1 \geq p \geq (2q)^{-1} \geq 1/2$ (call this region 2),
\begin{equation}\label{e:region2}
    \varepsilon_{Q}(G_{AND}^{p,q}) \leq \frac{1}{\sqrt{2}}\left[1 - 2\left(1-p\right)\left(1-q\right)\right].
\end{equation}
Note that the bounds \eqref{e:region1} and \eqref{e:region2} are strictly greater than $\varepsilon_{C} = 1/2$ for all values of $p$ and $q$ in the appropriate regions unless $p = q = 1/2$. Furthermore, there exist quantum strategies which attain these upper bounds. For region 1, Alice and Bob can share a maximally entangled state of four qubits (two for Alice and two for Bob) and choose measurement operators:
\begin{align*}
B_{00} &= X\otimes I,\hspace{0.35 in}  B_{10}= \left(\cos\beta X + \sin\beta Z\right)\otimes I,\\
B_{01} & = Y\otimes X, \hspace{0.35 in} B_{11} = -Y\otimes\left(\cos\beta X + \sin\beta Z\right),\\
A_{00} &= \left[p\left(\overline{B}_{00}+\overline{B}_{01}\right) + \left(1-p\right)\left(\overline{B}_{10}-\overline{B}_{11}\right)\right]/N_{00},\\
A_{01} &= \left[p\left(\overline{B}_{00}+\overline{B}_{01}\right) + \left(1-p\right)\left(-\overline{B}_{10}+\overline{B}_{11}\right)\right]/N_{10},\\
A_{10} &= \left[p\left(\overline{B}_{00}-\overline{B}_{01}\right) + \left(1-p\right)\left(\overline{B}_{10}+\overline{B}_{11}\right)\right]/N_{10},\\
A_{11} &= \left[p\left(-\overline{B}_{00}+\overline{B}_{01}\right) + \left(1-p\right)\left(\overline{B}_{10}+\overline{B}_{11}\right)\right]/N_{11},
\end{align*}
where
$\cos\beta = \frac{1}{2}\frac{(p^{2}+(1-p)^{2})(q^{2}+\left(1-q\right)^{2})}{p(1-p)(q^{2}+(1-q)^{2})}$,
$N_{00} = N_{10} = 2q\sqrt{\frac{p^{2}+(1-p)^{2}}{q^{2}+(1-q)^{2}}}$, and
$N_{01} = N_{11} = \frac{1-q}{q}N_{00}$.
For region 2, an optimal strategy corresponds to Alice and Bob sharing a maximally entangled state of two qubits and choosing measurement operators:
\begin{align*}
B_{00} &= B_{10} = X, \hspace{0.35 in} B_{01} = - B_{11}= Z,\\
A_{00} &= \left[p\left(\overline{B}_{00}+\overline{B}_{01}\right) + \left(1-p\right)\left(\overline{B}_{10}-\overline{B}_{11}\right)\right]/M_{00},\\
A_{01} &= \left[p\left(\overline{B}_{00}+\overline{B}_{01}\right) + \left(1-p\right)\left(-\overline{B}_{10}+\overline{B}_{11}\right)\right]/M_{10},\\
A_{10} &= \left[p\left(\overline{B}_{00}-\overline{B}_{01}\right) + \left(1-p\right)\left(\overline{B}_{10}+\overline{B}_{11}\right)\right]/M_{10},\\
A_{11} &= \left[p\left(-\overline{B}_{00}+\overline{B}_{01}\right) + \left(1-p\right)\left(\overline{B}_{10}+\overline{B}_{11}\right)\right]/M_{11},
\end{align*}
where
$M_{00} = M_{10} = \sqrt{2}$, and $M_{01} = M_{11} = \sqrt{2}(2p-1)$.
Thus, if Alice and Bob have some knowledge about the input bits $z_{i}$ (even an infinitesimal amount), quantum strategies do offer an advantage over classical strategies for computing the $2$-bit nonlocal AND function.

We now show that there exists a family of games corresponding to
 functions with input sizes of increasing length for which the same is
 true. Underlying the game is a Boolean function
 $g:\left\{0,1\right\}^{n}\rightarrow\left\{0,1\right\}$. We partition
 the inputs $z=z_1,\ldots,z_n$ in two groups $A$ and $B$, one for
 which Alice has some information, Pr$[z_i=x_i]=p_i, i\in A$, and the
 rest for which Bob has some information, Pr$[z_j=y_j]=p_j, j\in B$.
 For all $i$, $p_i \geq 1/2$. Each player thus has $n$ inputs
 $x_i,y_i$. The distribution $\pi$ over these inputs is specified as
 follows. The input $z$ is  distributed according to some distribution
 $P$. First pick $z$ according to $P$, 
 and then for $i\in B$
 choose $x_i=z_i$ and $y_i=0$ with probability $p_i$, and $x_i=1\oplus z_i$
 and $y_i =1$ with probability $1-p_i$.  Similarly for $j \in B$.
 Alice and Bob win the game iff $a\oplus b = g(x_1\oplus
 y_1,\ldots,x_n\oplus y_n)=f(x,y)$. 
% Note that when for all $i$
% $p_i=q_i=1/2$ the game satifies the no knowledge
% condition~\eqref{e:noknowledge}. 
 Note that the game satifies the no knowledge
 condition~\eqref{e:noknowledge} when
 $p_i=q_i=1/2$ for all $i$.
 We say that Alice or Bob have {\em
 some knowledge} of the game if not all $p_i$ and $q_i$ equal $1/2$.

Given any $n$-bit game $G_{1}=(f_{1},\pi_{1})$ and any $m$-bit game $G_{2}=(f_{2},\pi_{2})$, %one can
define their \textit{sum} to be the game
\begin{equation*}
    G_{1}\oplus G_{2} = \left(f_{1}\oplus f_{2}, \pi_{1}\times\pi_{2}\right).
\end{equation*}
In this game, the verifier selects pairs of strings $\left((x^{(1)},y^{(1)}),(x^{(2)},y^{(2)})\right)$
according to the product distribution $\pi_{1}\times\pi_{2}$.
%hmb shortened this a bit
Alice and Bob win the game if on input $x^{(1)}, x^{(2)}$ and $y^{(1)},
 y^{(2)}$, they output $a$ and $b$ such that $a\oplus b =
 f_1(x^{(1)},y^{(1)}) \oplus f_2(x^{(2)},y^{(2)})$.
 It was proved in \cite{CSUU08} that the quantum
 bias of a sum of games is simply the product of the individual
 biases. This multiplicativity makes it easy to compute the quantum bias of a sum of games if the individual biases are known. Unfortunately it does not hold for the classical bias, which is in general MAXSNP hard to compute \cite{AN04}.

Now 
%,for $k\in\mathbb{N}$ 
define the family of nonlocal distributed AND
games by taking the $k$-fold sum of the game $G^{\frac{1}{2},\frac{1}{2}}_{AND}$.
\begin{equation}
    G_{AND(k)} = \bigoplus_{j = 1}^{k}G^{\frac{1}{2},\frac{1}{2}}_{AND}. \label{e:ip2k}
\end{equation}
By construction these games satisfy the `no knowledge' condition
\eqref{e:noknowledge}, and hence there is no quantum advantage for any
of them. Indeed, it follows from~\cite{CSUU08}   that $\varepsilon_{C}(G_{AND(k)}) = \varepsilon_{Q}(G_{AND(k)}) = (1/2)^{k}$. As before, we can now consider allowing Alice and Bob some knowledge of the inputs, and define a family of perturbed games:
\begin{equation}
    G_{AND(k)}^{p,q} = G_{AND(k-1)} \oplus G_{AND}^{p,q}.
\end{equation}
In other words, this is the family of games formed by taking the sum
of $k-1$ copies of $G^{\frac{1}{2},\frac{1}{2}}_{AND}$ and a single
copy of $G_{AND}^{p,q}$. We now show that for all $k$, there is a
small region of $(p,q)$-space around $p = q = 1/2$ in which a quantum
advantage does exist. Fixing $q=1/2$, we allow $p$ to vary in the
interval $\left[1/2,1\right)$ (thus we remain in region 1). It follows
from~\cite{CSUU08} and equation~\eqref{e:region1} that the quantum bias is
\begin{equation}\label{e:quantumvalue}
    \varepsilon_{Q}(G_{AND(k)}^{p,1/2}) = \frac{1}{2^{k-1}\sqrt{2}}\sqrt{p^{2}+(1-p)^{2}}.
\end{equation}
Observe that the success probability of {\em any} classical strategy
will be a linear function in $p$, say $ap +b$, with $a$ and $b$ from a
finite  set of  values. Moreover the quantum success probability 
equation~\eqref{e:quantumvalue} is monotone increasing for $p\geq 1/2$,
and its derivative is $0$ for $p=1/2$. Since for $p=1/2$ the quantum
and the classical success probabilities are the same, and the
classical value is always less than or equal to the quantum value, it follows that
the line induced by the best classical protocol around $p=1/2$ is
horizontal and not increasing.
%hmb replaced this part with derivative argument.
%% which is a real analytic function for $p\in\left[0,1\right)$, but not a polynomial. Now, although the classical bias is not easily computable, it follows from \eqref{e:c_bias} and \eqref{e:pq} that $\varepsilon_{C}(G_{IP(2k)}^{p,1/2})$ must be a polynomial in $p$.  We also know that it must satisfy the trivial bound $\varepsilon_{C}\leq\varepsilon_{Q}$. Now, the set of points where a real analytic function $f(x)$ and a polynomial $g(x)$ are equal is the zero set of the analytic function $f-g$. However, in one real variable, if the zero set of a real analytic function has a limit point in a connected open set, it must be identically zero \cite{Kra82}. Since $\varepsilon_{C}(p)\neq\varepsilon_{Q}(p)$, the zero set of $\varepsilon_{C}-\varepsilon_{Q}$ cannot have a limit point in any connected open set around $p = 1/2$.
%% Since we know that the biases are equal at $p = q = 1/2$, there must
%% be a small region around this point where the quantum bias is strictly
%% greater than the classical bias.
Thus, an infinitesimal amount of
knowledge is sufficient for a quantum advantage to exist for this
family of games. Indeed, it is clear from the construction of this
example, that a quantum advantage also exists for any game $G =
G_{1}\oplus G_{AND}^{p,1/2}$, where $G_{1}$ has no quantum advantage
and $p>1/2$.

\textit{Discussion.} --- We have shown that perturbing a nonlocal computation game by allowing Alice and Bob an arbitrarily small amount of local knowledge can be enough to give a quantum advantage over classical strategies.

Let us now make a few remarks.  Firstly, as mentioned before, the bias of the game
$G_{AND}^{p,q}$ for $p=q=1$ is equivalent to the standard CHSH expression
(i.e. $\varepsilon_{C}=1/2$, $\varepsilon_{Q}=1/\sqrt{2}$). However the case $p=1$, $q=1/2$ is similar to the standard CHSH game, in that Alice has
complete knowledge of her bit (i.e. $x_{1}=z_{1}$).  However, the second bit $z_{2}$ is
completely unknown to Alice or Bob. Nonetheless, quantum protocols can do better than
classical ones, and achieve the same value as in the true CHSH case. The CHSH scenario is usually understood as concerning the situation where Alice and Bob
each have a bit locally and they output a bit: the output for the task being the XOR of the output bits.
However the above observation shows that in fact equal success can be achieved in this task if, say,
Alice knows one of the input bits, but neither Alice nor Bob have any knowledge of the other
input bit; Bob can have no local information at all.  

In this Letter we have focused on allowing Alice and Bob a particular kind of knowledge of
the inputs of $f$. In fact, more general perturbations can be considered by relaxing the
`no knowledge' condition \eqref{e:noknowledge}
completely, and allowing the probability distribution
$\pi(x,y)$ to be arbitrary. Then, given a game $G =(f,\pi)$ for which
$\varepsilon_{C} = \varepsilon_{Q}$, does there always exist a perturbed game
$G^{\prime} = (f,\pi^{\prime})$ which is arbitrarily close to $G$
(in some appropriate distance measure) for which a quantum advantage does exist?  We do
not have a general characterization of the games for which this is true, but we end with an
example. Consider the game $G_{AND}^{M} = (f,\pi)$ where $f$ is the $2$-bit
nonlocal AND function given by \eqref{e:nland} and where $\pi(x,y) = (1/136)M_{xy}$ ,
where the matrix $M$ is the $4\times 4$ magic square:
\begin{equation}
M =
\left[
\begin{array}{cccc}
     4 &  14  &   15 & 1 \\
     9 &  7 &  6 & 12\\
     5  &  11  &   10 & 8 \\
     16  &  2 &  3 & 13 \end{array}
\right].
\end{equation}
Exhaustive search over all classical strategies shows that 
%It can be shown using exhaustive search 
%over all classical strategies that
$\varepsilon_{C}(G_{AND}^{M}) = 1/2$. On the other hand, using semidefinite
programming (e.g. using the Matlab packages SeDuMi\cite{SeDuMi} and YALMIP \cite{yalmip}), it follows  that $\varepsilon_{Q}(G_{AND}^{M}) \geq
0.5911$. This is interesting because the marginal distributions
$\pi(x) = \sum_{y}\pi(x,y)$ and $\pi(y) = \sum_{x}{\pi(x,y)}$ are
identical to the marginals of the game $G_{AND}$, for which no quantum
advantage exists. In both cases, the two marginals (which correspond
to the marginal probabilities that Alice and Bob receive strings $x$
and $y$ respectively) are equal to $1/4$, independent of $x$ and
$y$. So this game is locally  indistinguishable to Alice and Bob from $G_{AND}$, and yet a quantum advantage exists for $G_{AND}^{M}$.  It would be interesting to know, for instance, whether it is possible to have a game where the marginals are random but the quantum value is equal to that of the CHSH game. 

We thank M.~van den Berg, T.~Cubitt, T.~Lawson and J.~Sharam for helpful
conversations. We gratefully acknowledge support for this work from:
the Dorothy Hodgkin Foundation [JA]; the NWO Vici-project 639.023.302
and BRICKS project AFM1 [HB], the UK EPSRC through the
QIP-IRC and the University of Bristol for a Research
Fellowship [NL];  the EU
project QAP [HB and NL].

%%%%%%%%%%%%%%%%%%%%%%%%%%%%%%%%%%%%%%%%%%%%%%%%%%%%%%%%%%%%%%%%%%%%%%%%%%%%%%%%%%%%%%%%%%%%%%%%


\begin{thebibliography}{13}
\expandafter\ifx\csname natexlab\endcsname\relax\def\natexlab#1{#1}\fi
\expandafter\ifx\csname bibnamefont\endcsname\relax
  \def\bibnamefont#1{#1}\fi
\expandafter\ifx\csname bibfnamefont\endcsname\relax
  \def\bibfnamefont#1{#1}\fi
\expandafter\ifx\csname citenamefont\endcsname\relax
  \def\citenamefont#1{#1}\fi
\expandafter\ifx\csname url\endcsname\relax
  \def\url#1{\texttt{#1}}\fi
\expandafter\ifx\csname urlprefix\endcsname\relax\def\urlprefix{URL }\fi
\providecommand{\bibinfo}[2]{#2}
\providecommand{\eprint}[2][]{\url{#2}}

\bibitem[{\citenamefont{Ekert}(1991)}]{Eke91}
\bibinfo{author}{\bibfnamefont{A.}~\bibnamefont{Ekert}},
  \bibinfo{journal}{Phys. Rev. Lett.} \textbf{\bibinfo{volume}{67}},
  \bibinfo{pages}{661} (\bibinfo{year}{1991}).

\bibitem[{\citenamefont{Bennett et~al.}(1993)\citenamefont{Bennett, Brassard,
  Cr\'{e}peau, Jozsa, Peres, and Wootters}}]{BBCJ+93}
\bibinfo{author}{\bibfnamefont{C.}~\bibnamefont{Bennett}},
  \bibinfo{author}{\bibfnamefont{G.}~\bibnamefont{Brassard}},
  \bibinfo{author}{\bibfnamefont{C.}~\bibnamefont{Cr\'{e}peau}},
  \bibinfo{author}{\bibfnamefont{R.}~\bibnamefont{Jozsa}},
  \bibinfo{author}{\bibfnamefont{A.}~\bibnamefont{Peres}}, \bibnamefont{and}
  \bibinfo{author}{\bibfnamefont{W.}~\bibnamefont{Wootters}},
  \bibinfo{journal}{Phys. Rev. Lett.} \textbf{\bibinfo{volume}{70}},
  \bibinfo{pages}{1895} (\bibinfo{year}{1993}).

\bibitem[{\citenamefont{Bennett and Wiesner}(1992)}]{BW92}
\bibinfo{author}{\bibfnamefont{C.}~\bibnamefont{Bennett}} \bibnamefont{and}
  \bibinfo{author}{\bibfnamefont{S.}~\bibnamefont{Wiesner}},
  \bibinfo{journal}{Phys. Rev. Lett.} \textbf{\bibinfo{volume}{69}},
  \bibinfo{pages}{2881} (\bibinfo{year}{1992}).

\bibitem[{\citenamefont{Cleve and Buhrman}(1997)}]{BC97}
\bibinfo{author}{\bibfnamefont{R.}~\bibnamefont{Cleve}} \bibnamefont{and}
  \bibinfo{author}{\bibfnamefont{H.}~\bibnamefont{Buhrman}},
  \bibinfo{journal}{Phys. Rev. A} \textbf{\bibinfo{volume}{56}},
  \bibinfo{pages}{1201} (\bibinfo{year}{1997}).

\bibitem[{\citenamefont{de~Wolf}(2002)}]{Wol02}
\bibinfo{author}{\bibfnamefont{R.}~\bibnamefont{de~Wolf}},
  \bibinfo{journal}{Theor. Comp. Science} \textbf{\bibinfo{volume}{287}},
  \bibinfo{pages}{337} (\bibinfo{year}{2002}).

\bibitem[{\citenamefont{Clauser et~al.}(1969)\citenamefont{Clauser, Horne,
  Shimony, and Holt}}]{CHSH69}
\bibinfo{author}{\bibfnamefont{J.}~\bibnamefont{Clauser}},
  \bibinfo{author}{\bibfnamefont{M.}~\bibnamefont{Horne}},
  \bibinfo{author}{\bibfnamefont{A.}~\bibnamefont{Shimony}}, \bibnamefont{and}
  \bibinfo{author}{\bibfnamefont{R.}~\bibnamefont{Holt}},
  \bibinfo{journal}{Phys. Rev. Lett.} \textbf{\bibinfo{volume}{23}},
  \bibinfo{pages}{880} (\bibinfo{year}{1969}).

\bibitem[{\citenamefont{Linden et~al.}(2007)\citenamefont{Linden, Popescu,
  Short, and Winter}}]{LPSW07}
\bibinfo{author}{\bibfnamefont{N.}~\bibnamefont{Linden}},
  \bibinfo{author}{\bibfnamefont{S.}~\bibnamefont{Popescu}},
  \bibinfo{author}{\bibfnamefont{A.}~\bibnamefont{Short}}, \bibnamefont{and}
  \bibinfo{author}{\bibfnamefont{A.}~\bibnamefont{Winter}},
  \bibinfo{journal}{Phys. Rev. Lett.} \textbf{\bibinfo{volume}{99}},
  \bibinfo{pages}{180502} (\bibinfo{year}{2007}).

\bibitem[{\citenamefont{Brassard et~al.}(2006)\citenamefont{Brassard, Buhrman,
  Linden, M\'{e}thot, Tapp, and Unger}}]{BBLM+06}
\bibinfo{author}{\bibfnamefont{G.}~\bibnamefont{Brassard}},
  \bibinfo{author}{\bibfnamefont{H.}~\bibnamefont{Buhrman}},
  \bibinfo{author}{\bibfnamefont{N.}~\bibnamefont{Linden}},
  \bibinfo{author}{\bibfnamefont{A.}~\bibnamefont{M\'{e}thot}},
  \bibinfo{author}{\bibfnamefont{A.}~\bibnamefont{Tapp}}, \bibnamefont{and}
  \bibinfo{author}{\bibfnamefont{F.}~\bibnamefont{Unger}},
  \bibinfo{journal}{Phys. Rev. Lett.} \textbf{\bibinfo{volume}{96}},
  \bibinfo{pages}{250401} (\bibinfo{year}{2006}).

\bibitem[{\citenamefont{Cleve et~al.}(2004)\citenamefont{Cleve, Hoyer, Toner,
  and Watrous}}]{CHTW04}
\bibinfo{author}{\bibfnamefont{R.}~\bibnamefont{Cleve}},
  \bibinfo{author}{\bibfnamefont{P.}~\bibnamefont{Hoyer}},
  \bibinfo{author}{\bibfnamefont{B.}~\bibnamefont{Toner}}, \bibnamefont{and}
  \bibinfo{author}{\bibfnamefont{J.}~\bibnamefont{Watrous}},
  \bibinfo{journal}{19th Annual IEEE Conference on Computational Complexity} p.
  \bibinfo{pages}{236} (\bibinfo{year}{2004}).

\bibitem[{\citenamefont{Cleve et~al.}(2008)\citenamefont{Cleve, Slofstra,
  Unger, and Upadhyay}}]{CSUU08}
\bibinfo{author}{\bibfnamefont{R.}~\bibnamefont{Cleve}},
  \bibinfo{author}{\bibfnamefont{W.}~\bibnamefont{Slofstra}},
  \bibinfo{author}{\bibfnamefont{F.}~\bibnamefont{Unger}}, \bibnamefont{and}
  \bibinfo{author}{\bibfnamefont{S.}~\bibnamefont{Upadhyay}},
  \bibinfo{journal}{Computational Complexity} \textbf{\bibinfo{volume}{17}},
  \bibinfo{pages}{282} (\bibinfo{year}{2008}).

\bibitem[{\citenamefont{Alon and Naor}(2004)}]{AN04}
\bibinfo{author}{\bibfnamefont{N.}~\bibnamefont{Alon}} \bibnamefont{and}
  \bibinfo{author}{\bibfnamefont{A.}~\bibnamefont{Naor}},
  \bibinfo{journal}{Proceedings of the Thirty-Sixth Annual ACM Symposium on
  Theory of Computing}  (\bibinfo{year}{2004}).

\bibitem[{\citenamefont{Sturm}(1999)}]{SeDuMi}
\bibinfo{author}{\bibfnamefont{J.}~\bibnamefont{Sturm}},
  \bibinfo{journal}{Optimization Methods for Software}
  \textbf{\bibinfo{volume}{11}}, \bibinfo{pages}{625} (\bibinfo{year}{1999}).

\bibitem[{\citenamefont{Lofberg}(2004)}]{yalmip}
\bibinfo{author}{\bibfnamefont{J.}~\bibnamefont{Lofberg}},
  \bibinfo{journal}{Proceedings of the CACSD Conference, Taipei, Taiwan}
  (\bibinfo{year}{2004}).

\end{thebibliography}
\end{document}